\documentclass{moriond}

\def\be{\begin{equation}}
\def\ee{\end{equation}}
\def\bea{\begin{eqnarray}}
\def\eea{\end{eqnarray}}

\newcommand{\Photo}{\includegraphics[height=35mm, angle=-90]{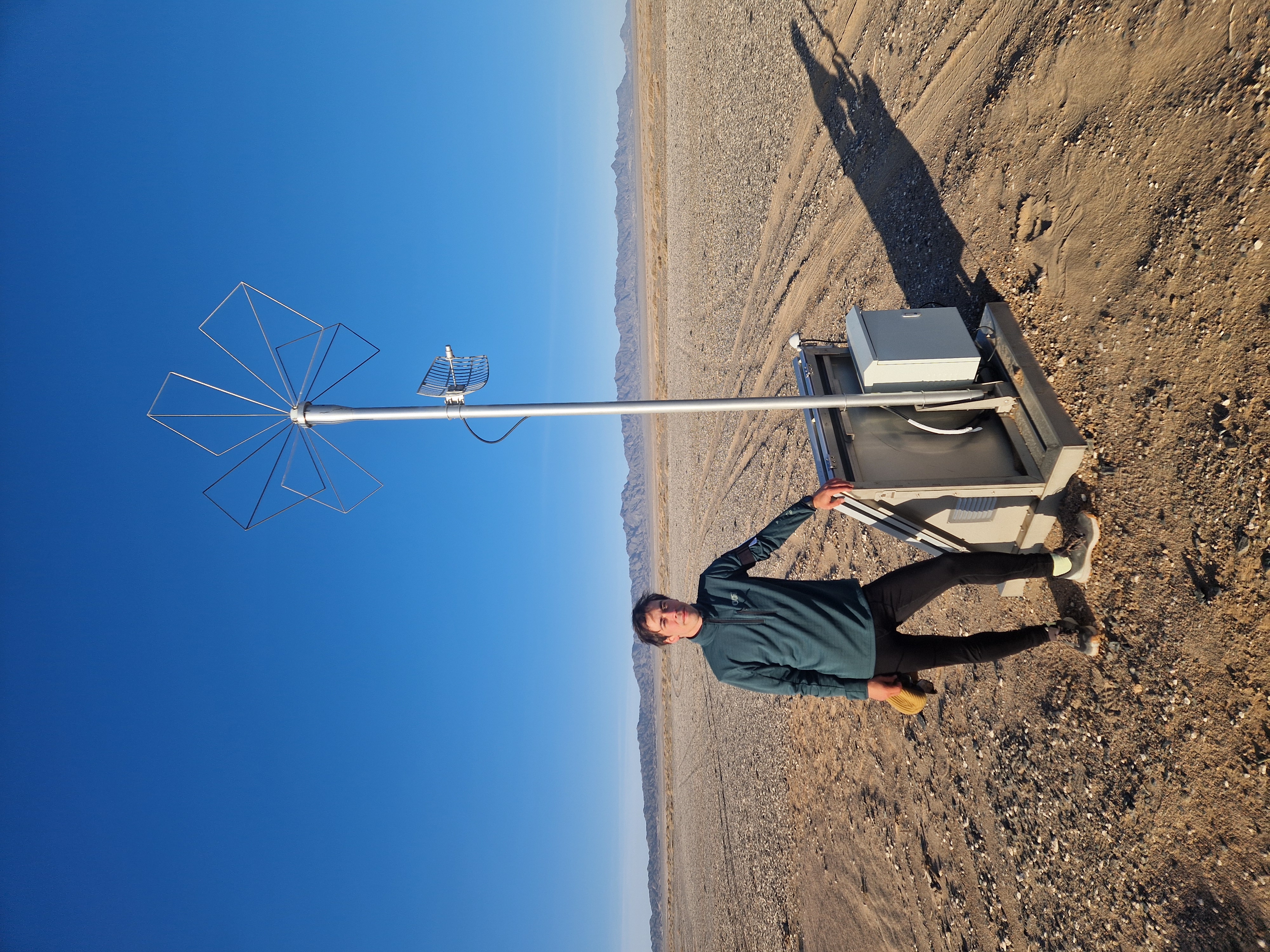}}
\usepackage[utf8]{inputenc}
\usepackage{enumitem}
\usepackage{amsmath}
\usepackage{svg}
\usepackage{multicol}
\usepackage{float}

\begin{document}
\vspace*{4cm}
\title{Radio signal generation in milliseconds: enabling multi-parameter reconstruction of ultra-high-energy cosmic rays}

\author{ A. Ferriere for the GRAND Collaboration}

\address{Sorbonne Université, CNRS, Laboratoire de Physique Nucléaire et des Hautes Energies (LPNHE)\\
4 Pl. Jussieu, 75005 Paris, France\\
Université Paris-Saclay, CEA, List, F-91120, Palaiseau, France}

\maketitle\abstracts{
In recent years, radio detection of ultra-high-energy cosmic rays (UHECRs), with energies above $10^{18}$ eV, has become an established technique. The radio emissions can be simulated with high accuracy using Monte Carlo codes such as ZHAireS and CoREAS. These simulations are essential but are computationally intensive. In this work, we present a machine-learning-based emulator that reproduces radio signal simulations with high accuracy in milliseconds rather than hours. Primary particle properties can then be reconstructed by comparing measured signals to emulated traces using a Markov Chain Monte Carlo approach. Using ZHAireS simulations carried out over the GRANDProto300 experiment layout, the method achieves an 8.9\% resolution on electromagnetic energy and a 0.08° angular resolution, matching state-of-the-art reconstruction performance. Finally, we apply the method on real data, successfully reconstructing cosmic-ray candidates detected by the GP300 prototype.}

\section*{Introduction}

Upon entering the atmosphere, ultra high energy cosmic rays (UHECRs) interact with atoms in the air and produce extensive air showers (EAS) of secondary particles. The produced electrons and positrons emit radio signals, primarily due to the influence of the Earth's magnetic field (the geomagnetic effect~\cite{geom}). Monte Carlo simulators such as ZHAireS and CoREAS~\cite{ZHAireS,CoREAS} simulate the shower and its radio emissions with high accuracy but at a high computational cost, typically several hours. This makes it difficult to use these simulations for reconstructing the properties of cosmic rays with a Bayesian approach which often requires sampling the input space. In this work, we present a machine learning based emulator that is trained on ZHAireS simulations and predicts in milliseconds the measured signal, enabling reconstruction. We first describe the emulator and its performance on simulations, then how to use it for reconstructing cosmic rays, with an application to GP300 data.

\section{Machine-learning emulator for radio signals generation}\label{sec:emulator}
The emulator is a machine learning model that takes as input the parameters of the shower, namely the positions of the antennas, the position of the shower maximum $X_{\max}$, the electromagnetic energy and the direction of the shower. We assume that no other parameter affects the radio signal. 
\subsection{Data Parametrisation}
{\bf Shower parametrisation:} Rather than using directly the position of the shower maximum and the position of the antenna, we prefer using relative positioning of the antenna with respect to the shower maximum. We express this relative positioning in a coordinate system defined by the shower axis ($\vec{k}$) and the magnetic field ($\vec{B}$): $(\vec{X}_{\max},\ (\vec{k},\, \vec{k}\times\vec{B},\, \vec{k}\times\vec{k}\times\vec{B}))$. It is similar to the parametrisation used for the angular distribution function (ADF)~\cite{adf}.

The list of inputs to characterise the signal from an event received by an antenna is:

\begin{itemize}[noitemsep]
    \begin{multicols}{2}
        \item Zenith angle $\theta$
        \item Azimuth angle $\phi$, ($\cos\phi$ and $\sin\phi$)
        \item Off-axis angle $\omega_i$ 
        \item Polar angle $\eta_i$, ($\cos\eta_i$ and $\sin\eta_i$)
        \item Antenna distance to $X_{\max}$, $l_i$
        \item Electromagnetic energy of the shower $E$
    \end{multicols}\vspace{-.3cm}
    \item z-component of $X_{\max}$ and effective refractive index $n_{{\rm eff}, i}$ to keep information about the atmospheric depth of the shower maximum. 
    \end{itemize}
{\bf Radio signal parametrisation:} The emulator outputs the radio signal that would be measured at the antenna positions. To reduce the dimensionality of the output space, we use a parametrisation of the signal. For a given antenna $i$ at position $\vec{X}_i$, the signal is polarised along $\vec{k} \times \vec{B}$ for the dominant geomagnetic emissions. As a simplification, we only consider this polarisation. The signal can then be described by a few parameters in the frequency domain. If, $S(f) = |S(f)| e^{i\Phi(f)}$ is the Fourier transform of the signal, we use the following parametrisation of the signal, inspired by similar approaches~\cite{param}:
\begin{equation}
\begin{aligned}
    |S(f)| &= \exp\left(a + b(f-f_0) + c(f-f_0)^2 \right) \\
    \Phi(f) &= \Phi_0 + \Phi_{\rm p} (f-f_0) + \Phi_{\rm q} (f-f_0)^2
\end{aligned}
\vspace{-0.1cm}
\end{equation}
where $f_0$ is a reference frequency (set to $30$ MHz). 5 parameters are needed to describe the signal, $a$, $b$, $c$ for the amplitude and $\Phi_p$ and $\Phi_q$ for the phase. $\Phi_0$ is set so that the phase is $\pi$ at 0 MHz, to force the signal to be real valued in the time domain. It allows us to reduce the output space dimensionality from 1024 time bins × 3 polarization channels to 5 parameters. This parametrisation holds for $f < f_{\rm thin}$, where $f_{\rm thin}$ is automatically determined. Above this frequency, the signal is dominated by simulation noise from thinning as particle packets emit coherent radiation while individual particles would not. For data quality, we retain only signals with $f_{\rm thin} > 80$ MHz to ensure that the signal is not dominated by simulation noise.

% where $f_0$ is a reference frequency (here $30$ MHz). 5 parameters are needed to describe the signal, $a$, $b$, $c$ for the amplitude and $\Phi_p$ and $\Phi_q$ for the phase. $\Phi_0$ is set so that the phase is $\pi$ at 0 MHz, to force the signal to be real valued in the time domain. This parametrisation holds for $f < f_{\rm thin}$, frequency above which the signal is dominated by simulation noise from thinning.}

\subsection{Model architecture, training and performance}
The model is a feedforward neural network with 4 hidden layers of 300 neurons each. The model is trained on a set of 15000 ZHAireS simulations of iron and proton showers at the GRANDProto300 (GP300) site in the Gansu province of China to predict the parameters of the electric field signal. The loss function is the mean squared error between the predicted parameters and the true parameters. 
\begin{figure}
    \centering
    \includegraphics[width=0.75\textwidth, trim={0 0 0 175pt}, clip]{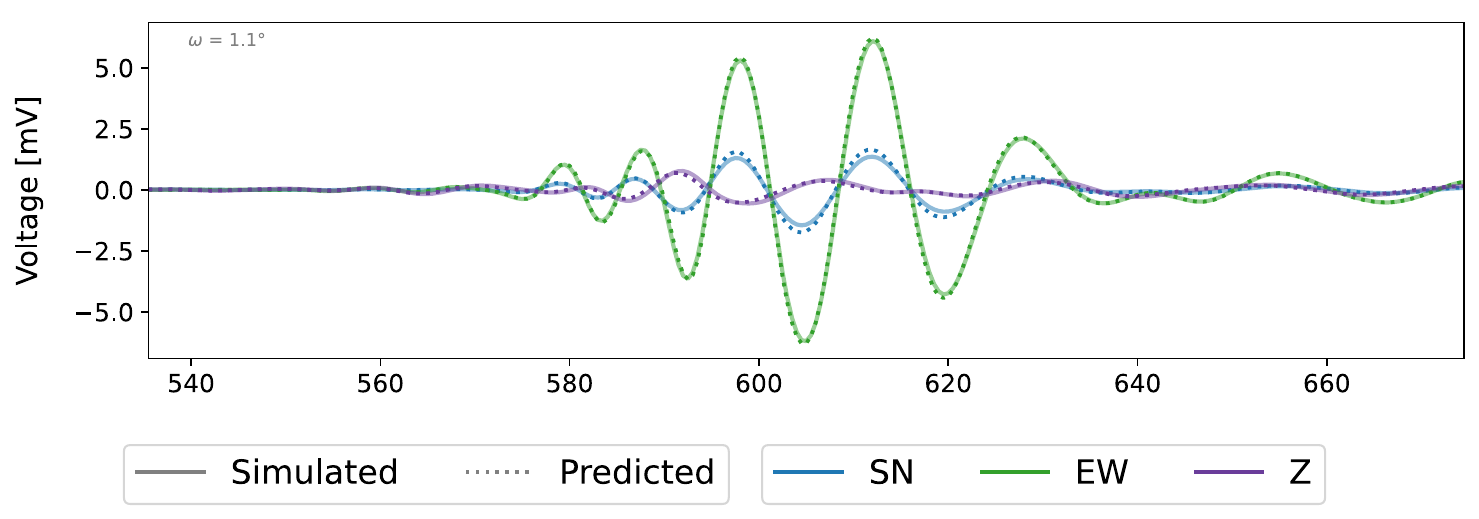}
    \caption{Comparison of the voltage traces predicted by the emulator (dotted) and simulated by ZHAireS (solid) for a given event.}\label{fig:emulator_example}\vspace{-0.2cm}
\end{figure}
At inference, we can convert the predicted electric field to voltage using a modeling of the antenna response~\cite{GNN}. This full conversion runs in 6 milliseconds on average for an event with 20 antennas on a personal computer. Such a voltage prediction, compared to the ZHAireS simulated one, can be seen in Fig.~\ref{fig:emulator_example}.
On more validation events, we can compare the predicted voltage amplitude and fluence with the simulated one for a set of validation simulations. The achieved relative error is 5.2\% on amplitude and 8.2\% on fluence, smaller than the error between different ZHAireS and CoREAS~\cite{JaimeICRC}.

\section{Bayesian Parameter Reconstruction}
The emulator can be used to reconstruct the properties of cosmic rays by comparing the measured signals to emulated traces. 

Related work uses Information Field Theory~\cite{StrahnzIFT} for shower reconstruction. We pursue a complementary approach: rapid Bayesian reconstruction via MCMC enabled by the emulator.

\subsection{Probabilistic modeling}
The goal of the reconstruction is to find $p(\theta, \phi, E, X_{\max} | \{V_i\})$, the posterior distribution of the shower parameters given the measured voltage traces on the antennas. Using Bayes' theorem, we can compute it from the likelihood and the prior distributions.

The likelihood is defined as the product of the likelihood of the voltage traces and the likelihood of the timing of the signal defined as follows: 

\begin{equation}
    \begin{aligned}
        p(\{V_i\} | \theta, \phi, E, X_{\max}) &= \mathcal{N}\left(\{V_i^{\rm emu}\},\ \sigma_n^2 + \sigma_s^2 \{V_i^{\rm emu}\}^2\right) \\
        p(\{t_i\} | \theta, \phi, E, X_{\max}) &= \mathcal{N}\left(\left\{\frac{l_i n_{{\rm eff}, i}}{c}\right\}, \sigma_t^2\right)
    \end{aligned}
    \vspace{-0.2cm}
\end{equation}
where $\{V_i^{\rm emu}\}$ are the emulated voltage traces, $\{t_i\}$ the times of arrival of the traces (time of maximum Hilbert amplitude), $\sigma_s$ is the standard deviation of the error of the emulator and of calibration,  $\sigma_n$ is the standard deviation of the noise on the measurement (galactic and thermal noise) and $\sigma_t$ is the standard deviation of the time jitter and of the error of the SWF model.

For now, the prior encodes two constraints: (1) the correlation between zenith angle and shower maximum depth via a polynomial fit to the training data for the mean and the standard deviation, and (2) zero probability for parameter combinations outside the distribution of the parameters in the training set. These priors help constrain the parameter space to physically realistic values based on the training data distribution.

\subsection{Reconstruction performance}
The reconstruction using a Monte Carlo Markov Chain approach was performed on a set of 1238 simulations of voltage signal with realistic noise and triggering conditions
% , the results are shown in Fig.~\ref{fig:reconstruction_performance}. 
We achieve an 8.9\% resolution on electromagnetic energy and a 0.08° angular resolution with calibrated uncertainties, matching state-of-the-art reconstruction performance. A conversion of the position of $X_{\max}$ to the atmospheric depth of the shower maximum gave a preliminary resolution of 81 g/cm$^2$.

Another method was explored where we compare the maximum amplitude of the signal rather than the full voltage traces, yielding a 0.05° angular resolution, a 12.5\% energy resolution, and a 104 g/cm$^2$ resolution on the atmospheric depth. This method is analogous to the ADF method as it considers the maximum amplitude of the signal.

\section{Application to GP300 cosmic ray candidates}
\begin{figure}[H]
\centering
\includegraphics[width=0.38\textwidth]{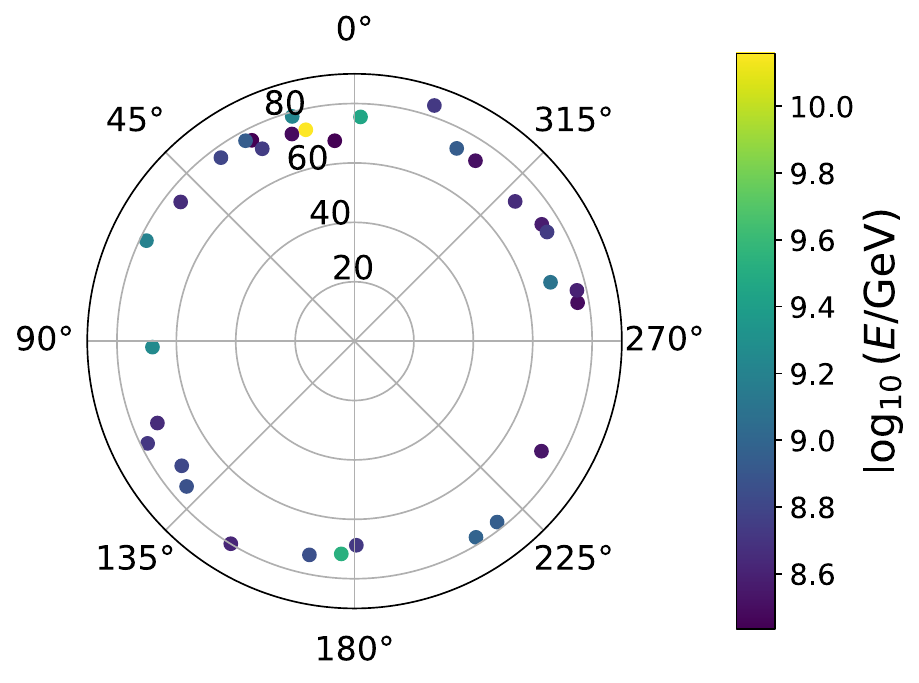}
\includegraphics[width=0.38\textwidth]{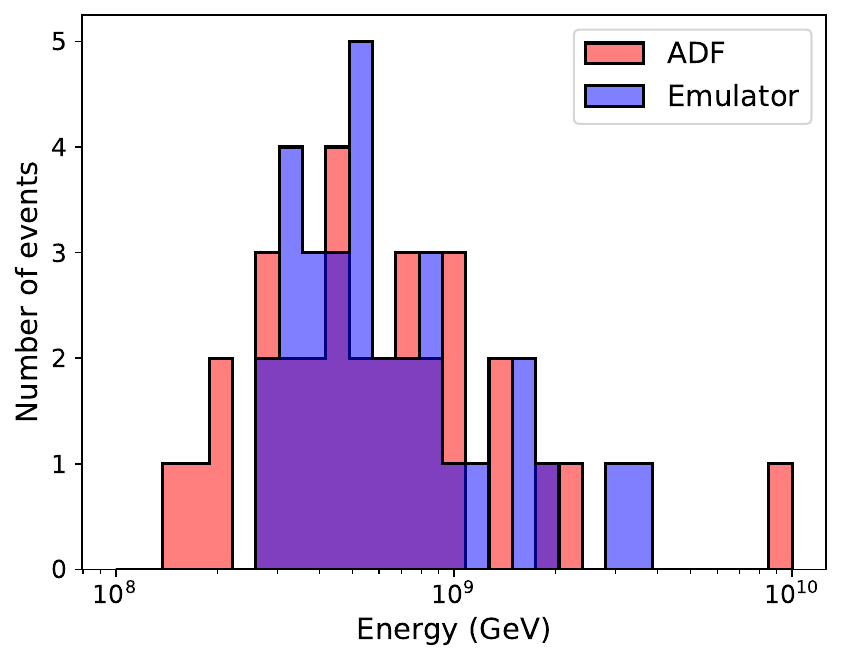}
    \caption{Reconstruction of the arrival directions (left) and energy distribution (right) compared with the ADF reconstruction. The GP300 prototype is in its commissioning phase; these results serve as proof of concept.}\label{fig:GP300_reconstruction}
\end{figure}

A set of 51 cosmic-ray candidates was selected from the data of the GP300 prototype~\cite{Jolan}. Among these, 32 were well reconstructed by the previously described method. We can construct an energy spectrum consistent with the one reconstructed by the ADF, and very similar reconstructed directions as shown by Fig.~\ref{fig:GP300_reconstruction}. 
\vspace{-0.2cm}
As the efficiency and purity of the cosmic-ray detection process remain to be fully characterized, these preliminary results demonstrate the validity of the emulator-based reconstruction approach on real data.

\vspace{-0.1cm}
\section*{Conclusion}
In this work, we presented a machine learning based emulator that can reproduce radio signal simulations with high accuracy in milliseconds rather than hours. This emulator is trained on processed ZHAireS simulations of iron and proton showers at the GP300 site and achieve a relative error of 5.2\% on the voltage amplitude, on the order of the discrepancy between ZHAireS and CoREAS.
We then used this emulator to reconstruct the properties of cosmic rays using a Markov Chain Monte Carlo approach, defining a likelihood that accounts for the signal on the antennas and the timing of the signal on the different antennas. This achieves a theoretical resolution of 8.9\% on electromagnetic energy and a 0.08° angular resolution on the GP300 layout. 

Finally, we applied the method on real data, reconstructing both the energy and arrival directions of 32 cosmic ray candidates detected by the GP300 prototype, with results consistent with the ones obtained by the angular distribution function method.

\footnotesize
\section*{Acknowledgments}
The author thanks J. Alvarez-Muñiz and IGFAE for the hospitality and support as well as O. Martineau-Huynh and A. Benoit-Lévy for their help. Simulations were performed using the computing resources at the IN2P3 Computing Centre (Lyon, France), a partnership between CNRS/IN2P3 and CEA/DSM/Irfu.\vspace{-.1cm}
\section*{References}
\bibliography{moriond}

%%% manually generated bibliography
%\begin{thebibliography}{99}
%\bibitem{ja}C Jarlskog in {\em CP Violation}, ed. C Jarlskog
%(World Scientific, Singapore, 1988).
%\bibitem{ma}L. Maiani, \Journal{\PLB}{62}{183}{1976}.
%\bibitem{bu}J.D. Bjorken and I. Dunietz, \Journal{\PRD}{36}{2109}{1987}.
%\bibitem{bd}C.D. Buchanan {\it et al}, \Journal{\PRD}{45}{4088}{1992}.
%\end{thebibliography}

\end{document}